Decreased aneurysmal subarachnoid hemorrhage incidence rate in elderly population than in middle aged population: a retrospective analysis of 8,144 cases in Mainland China

Yì Xiáng J. Wáng[1], Lihong Zhang[2], Lin Zhao[2], Jian He[3], Xian-Jun Zeng[4], Heng Liu[5], Yun-jun Yang[6], Shang-Wei Ding[7], Zhong-Fei Xu[8], Yong-Min He[10], Lin Yang[11], Lan Sun[12], Ke-jie Mu[12], Bai-Song Wang[12], Xiao-Hong Xu[13], Zhong-You Ji[14], Jian-hua Liu[15], Jin-Zhou Fang[16], Rui Hou[17], Feng Fan[18], Guang Ming Peng[19], Sheng-Hong Ju[20].

Correspondence to:  Dr Yì Xiáng Wáng, Department of Imaging and Interventional Radiology, The Chinese University of Hong Kong, Shatin, New Territories, Hong Kong SAR
yixiang_wang@cuhk.edu.hk

[1] Department of Imaging and Interventional Radiology, The Chinese University of Hong Kong, Shatin, New Territories, Hong Kong SAR

[2] Hebei Medical University No. 2 Hospital, Shijiazhuang, Hebei Province, China.

[3] Nanjing University Nanjing Drum Tower Hospital, Nanjing, Jiangsu Province, China.

[4] Nanchang University No.1 Hospital, Nanchang, Jiangxi Province, China.

[5] Zunyi Medical University Hospital, Zunyi, Guizhou Province, China.

[6] Wenzhou Medical University No. 1 Hospital, Wenzhou, Zhejiang Province, China.

[7] Dongguan People's Hospital, Teaching Hospital of Southern Medical University, Dongguan, Guangdong Province, China.

[8] Taizhou Central People's Hospital & Teaching Hospital of Taizhou College, Jiaojiang, Zhejiang Province. China.

[9] Suzhou University No.1 Hospital, Suzhou, Jiangsu Province, China.

[10] North Sichuan Medical College Hospital, Nanchong, Sichuan Province, China.

[11] The People's Hospital of Bishan District, Chongqing, China.

[12] Shaoxing Hospital, Teaching Hospital of China Medical University, Keqiao, Zhejiang Province , China.

[13] Guangdong Medical University Hospital, Zhanjiang, Guangdong Province, China.

[14] Fujian Medical University Union Hospital, Fuzhou, Fujian Province, China.

[15] The first People's Hospital of Guangzhou City, Guangzhou, Guangdong Province, China.

[16] Hospital of Juhua Group Corporation, Quzhou, Zhejiang Province, China.

[17] The Central People's Hospital of Siping City, Siping, Jilin Province, China.

[18] Zhengzhou University No.1 Hospital, Zhengzhou, Henan Province, China.

[19] The General Hospital of Guangzhou Military Command, Guangzhou, Guangdong Province, China.

[20] Southeast University Zhongda Hospital, Nanjing, Jiangsu Province, China.

**Abstract**

**Purpose:** Rupture of an intracranial aneurysm is the most common cause of subarachnoid haemorrhage (SAH), which is a life-threatening acute cerebrovascular event that typically affects working-age people. This study aims to investigate the aneurysmal SAH incidence rate in elderly population than in middle aged population in China.

**Materials and methods:** Aneurysmal SAH cases were collected retrospectively from the archives of 21 hospitals in Mainland China. All the cases collected were from September 2016 and backward consecutively for a period of time up to 8 years. SAH was initially diagnosed by brain computed tomography, and CT angiography (CTA) or digital subtraction angiography (DSA) was followed and SAH was confirmed to be due to cerebral aneurysm. When for cases multiple bleeding occurred, the age of the first SAH was used in this study. The total incidence from all hospital at each age were summed together for females and males; then adjusted by the total population number at each age for females and males. The total population data was from the 2010 population census of the People's Republic of China.

**Results:** In total there were 8,144 cases, with 4,861 females and 3,283 males. Our analysis shows for both females and males the relative aneurysmal SAH rate started to decrease after around 65 years old. The males the relative aneurysmal SAH rate might have started to decrease after around 55 years old.

**Conclusion:** In contrast to previous reports, our data demonstrated a decreased aneurysmal subarachnoid hemorrhage incidence rate in elderly population than in middle aged population. Our data therefore support the hypothesis that aneurysms do not grow progressively once they form but probably either rupture or stabilize and that very elderly patients are at a reduced risk of rupture compared with patients who are younger with the same-sized aneurysms.

Rupture of an intracranial aneurysm is the most common cause of subarachnoid haemorrhage (SAH), which is a life-threatening acute cerebrovascular event that typically affects working-age people [1]. Current medical management options in patients with unruptured cerebral aneurysms are limited, consisting largely of smoking cessation, blood pressure control, and neurosurgical or endovascular interventions [2]. The exact prevalence of unruptured intracranial aneurysms (UIAs) is unknown, but at least one in 20 to 30 adults is likely to carry an asymptomatic UIA. Approximately one quarter of these UIAs rupture in a lifetime [3]. Intracranial aneurysms are increasingly found on cross-sectional imaging, including CT and MRI, done for reasons unrelated to the aneurysm. UIAs in Japanese patients may have a higher risk of rupture due to a genetic or habitual background [4]. The rate of SAH in the Japanese population is reported to be higher Cacausian populations, as high as 96 per 100,000 annually [4.Nevertheless, the prevalence of UCAs is not significantly different among the populations studied [5], which might indicate that UCAs in Japanese patients may rupture more frequently than they do in the Caucasian population. The prevalence of hypertension or smoking in Japanese adults is not higher than that of the US population [4].

Unlike Caucasian and Japanese populations, studies describing the epidemiology of cerebral aneurysm in the Chinese population have been published remain even more limited. Many studies on Caucasian and Japanese populations suggested that the risk for rupture of cerebral aneurysms rupture increases with age [6]. In the systemic review of Rinkel et al [5], it was estimated the percentage risk of rupture of per 100 patient-years (95% CI) was 3.45 (1.4-7.0) for 40-59 yrs old subjects, while 5.72 (3.4-9.0) for 60-79 yrs subjects. Morita et al [6] performed a systemic review on Japanese patients reported the percentage risk of rupture of per 100 patient-years (95% CI) was 2.27 (1.6–3.2) for < 60 years subjects, and 2.99 (2.3-3.8) for≥ 60 subjects. The purpose of this study was to investigate whether in Chinese population, aneurysmal subarachnoid hemorrhage incidence increases in elderly subjects, and the gender related difference. Despite the risk factors, including hypertension, cigarette smoking and alcohol use, are more common in men, aneurysmal SAH belongs to a few diseases which the incidence is higher in women than in men [7, 8]. This study aims to investigate the aneurysmal SAH incidence rate in elderly population than in middle aged population in China.

Materials and methods

Aneurysmal SAH cases were collected retrospectively from the archives of 21 hospitals in Mainland China. All the cases collected were from September 2016 and backward consecutively for a period of time up to 8 years. SAH was initially diagnosed by brain computed tomography, and CT angiography (CTA) or digital subtraction angiography (DSA) was followed and SAH was confirmed to be due to cerebral aneurysm. When for cases multiple bleeding occurred, the age of the first SAH was used in this study.

The patient data were from the following 18 hospitals in Mainland China: (1) Hebei Medical University No. 2 Hospital, Shijiazhuang, Hebei province (n=1693 cases); (2) Nanjing University Medical School Nanjing Drum Tower Hospital, Nanjing (n= 1678 cases), Jiangsu Province; (4) Nanchang University No.1 Hospital (n=862 cases); Nanchang, Jiangxi Province; (3) Zunyi Medical University Hospital, Zunyi, Guizhou Province (n=443 cases); (4) Wenzhou Medical University No. 1 Hospital, Wenzhou, Zhejiang Province (n=474 cases); (5) North Sichuan Medical College Hospital, Nanchong, Sichuan Province (n=699 cases); (6) Suzhou University No.1 Hospital, Suzhou, Jiangsu Province (n=383 cases); (7) The Central People's Hospital & Teaching Hospital of Taizhou College, Jiaojiang, Zhejiang Province (n=315 cases). (8) Capital Medical University Beijing Friendship Hospital (n=206 cases); (9) The People's Hospital of Bishan District, Chongqing, (n=220 cases); (10) Guangdong Medical University Hospital, Zhanjiang Guangdong province (n=244 cases); (11) Dongguan People's Hospital, Teaching Hospital of Southern Medical University, Dongguan, Guangdong Province (n=209 cases); (12) Zhengzhou University No.1 Hospital, Zhengzhou, Henan Province (n=165); (13) Shaoxing Hospital, Teaching Hospital of China Medical University, Keqiao, Zhejiang Province, (n=122); (14) The General Hospital of Guangzhou Military Command (n=98 cases); (13) Hospital of Juhua Group Corporation, Quzhou, Zhejiang Province (n=89 cases); (14) Guangzhou No. 1 People's Hospital, Guangzhou, Guangdong Province (n=48 cases). (15)The Central People's Hospital of Siping City, Siping, Jilin Province (n=52 cases); (16) Southeast University Zhongda Hospital, Nanjing (n=51 cases); (17) Fujian Medical University Union Hospital, Fuzhou, Fujian Province (n=45 cases); (18) The First Affiliated Hospital of Xi'an Jiao Tong University (n=48 cases).

Among these data, some have been reported analyzing another aspect of characteristics [7-11]. The gender-specific relative age related incidence has not been analyzed reported. Additionally, this study included 2,775 new cases.

The total incidence from all hospital at each age were summed together for females and males; then adjusted by the total population number at each age for females and males. The total population data was from the 2010 population census of the People's Republic of China [12]. There is apparent low population segment aged around then 49 years old. This is due to the faming period from 1958-62 and associated lower birth rate during that period [13-15]. As our data were retrospectively retrieved, the exact mean SAH incidence year could not be precisely determined. Additionally, the faming period affected different regions in China to variable extent, also the patient case sampling in this study was stratified. The perfect adjustment for this lower birth rate was not possible. We tested and found the mean SAH incidence year of the patients was four years earlier offers the most reasonable fit in SAH incidence rate curve shape.

Results

In total there were 8,144 cases, with 4,861 females and 3,283 males, their population adjusted relative aneurysmal SAH rate is shown in Fig 1A. The 2010 population census of the People's Republic of China is shown in Fig 1B [12]. Fig 1A shows for both females and males the relative aneurysmal SAH rate started to decrease after around 65 years old. The males the relative aneurysmal SAH rate might have started to decrease after around 55 years old.

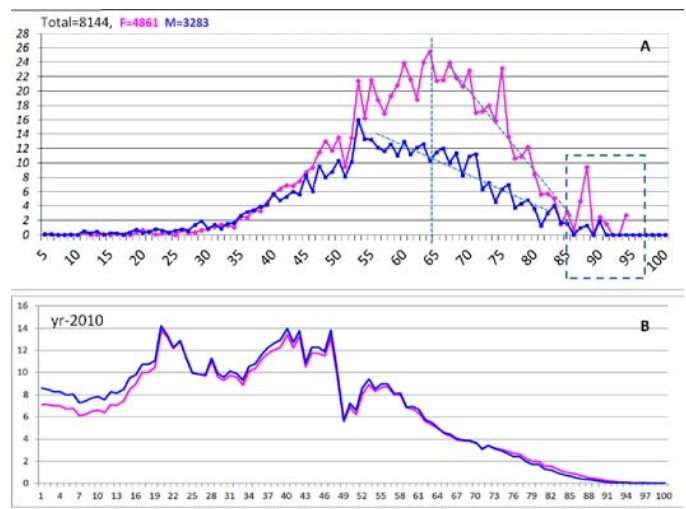

*Fig 1, A: Population adjusted relative aneurysmal SAH rate. X-axis number is arbitrary as the patient sampling was not stratified. Y-axis: years of age. The peaks after 85 yrs were caused by the small denominators. B: The 2010 population census of the People's Republic of China. X-axis: population number in millions. Y-axis: years of age in 2010.*

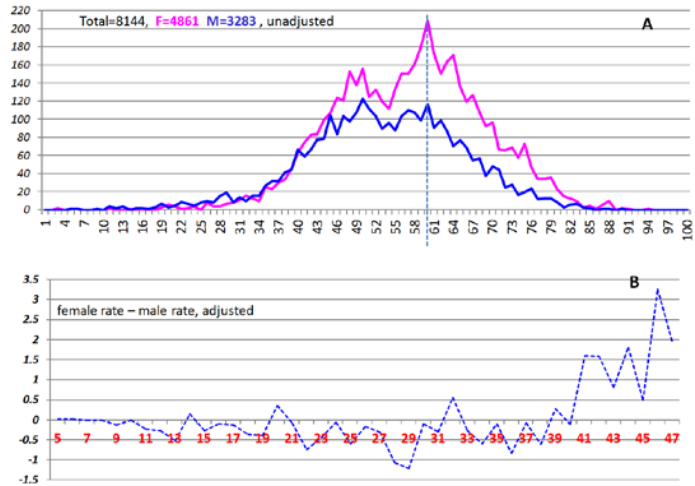

*Fig 2, A: Age specific distribution of the total 8,144 cases. X-axis: pooled aneurysmal SAH SAH incidence number. Y-axis: years of age. B: X-axis: population adjusted relative aneurysmal SAH rate difference between females and males. Y-axis: years of age (5 yrs to 47 yrs).*

The age specific distribution of the total 8,144 cases (4,861 females and 3,283 males) is shown in Fig 2A. Following the age drop centered around 49 yrs shown in Fig 1B, there was a relative lower aneurysmal SAH incidence period around 53 yrs. Fig 2B showed, the relative SAH incidence rate was higher in males than in females prior to 38 year old. After the population base for correction of the raw incidence, Stepwise Chi square test showed the relative SAH incidence rate was higher in males than in females before 38 year old. Women started to have a higher incidence of aneurysmal SAH than men after late thirties. This confirms our previous report [8].

The age specific distribution and population adjusted relative aneurysmal SAH rate of 1678 cases from Nanjing Drum Tower Hospital, Nanjing and 864 cases from Nanchang University No. 1 Hospital are shown in Fig 3 and Fig 4 respectively. These two data were randomly selected from the data pool and they demonstrated the same pattern shown in Fig 1 despite their small case number.

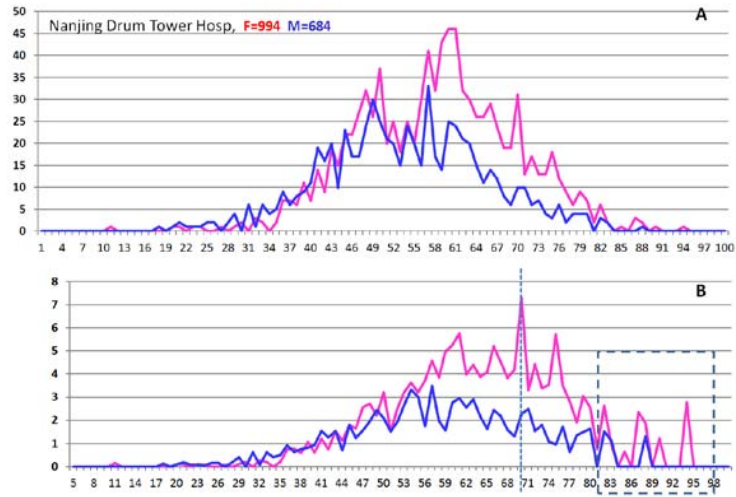

Fig 3, A: *Age specific distribution of the 1678 cases from Nanjing Drum Tower Hospital. B: Population adjusted relative aneurysmal SAH rate of the 1678 cases. X-axis number is arbitrary as the patient sampling was not stratified. Y-axis: years of age. The peaks after 80 yrs were caused by the small denominators.*

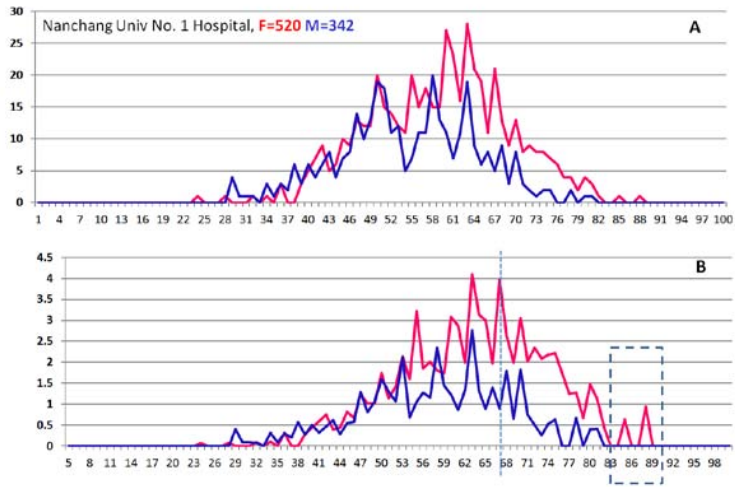

Fig 3, A: *Age specific distribution of the 864 cases from Nanchang University No. 1 Hospital. B: Population adjusted relative aneurysmal SAH rate of the 864 cases. X-axis number is arbitrary as the patient sampling was not stratified. Y-axis: years of age. The peaks after 83 yrs were caused by the small denominators.*

The case number from 60 years to 85 years is shown in Table 1. The absolute number of cases at each age decrease as the age increase.

Table: Patient number (female and male) at each age after 60 years old. Our data may suggest sufficient patient number to demonstrate these number decreases as the patient getting older.

| Age: yr | 60 | 61 | 62 | 63 | 64 | 65 | 66 | 67 | 68 | 69 |
|---------|-----|-----|-----|-----|-----|-----|-----|-----|-----|-----|
| Case no | 326 | 264 | 250 | 251 | 242 | 213 | 189 | 182 | 165 | 131 |
| Age: yr | 70 | 71 | 72 | 73 | 74 | 75 | 76 | 77 | 78 | 79 |
| Case no | 145 | 112 | 91 | 97 | 75 | 93 | 71 | 46 | 47 | 49 |
| Age: yr | 80 | 81 | 82 | 83 | 84 | 85 | | | | |
| Case no | 32 | 18 | 19 | 17 | 6 | 7 | | | | |

Discussion

Despite recent improvements in surgical and medical management of aneurysmal SAH, the overall mortality rate in this disease remains high [2]. The high mortality and morbidity rates are attributed mainly to brain damage caused by a severe initial hemorrhage, early re-bleeding, and delayed cerebral ischemia. During the last two decades, detection of unruptured intracranial aneurysms (UIAs) has increased because of new and improving diagnostic imaging technology, particularly magnetic resonance angiography (MRA), and computerized tomography angiography (CTA). When UIA is detected, deciding on the optimal treatment strategy requires careful assessment of each individual's treatment-related morbidity and life expectancy and knowledge of the natural history of this disease. Objective assessment of the risks of location-specific and size-specific contemporary endovascular and surgical management options is needed.

The risk of aneurysmal rupture without any intervention should be compared with the risks of endovascular treatment or surgical clipping. Treatable factors include cigarette smoking [16, 17] and hypertension [16, 18,]. Heavy alcohol use increases the risk of subarachnoid haemorrhage [17, 19]. Large aneurysms and those located in the posterior circulation are known to be associated with a high risk of rupture [20]. Prior history of SAH has been suggested to increase risk of rupture of an unruptured aneurysm [2]. Surgical or interventional treatment should also be in patients with a strong family history of subarachnoid haemorrhage, a daughter sac, or aneurysms at potentially high-risk locations, including the posterior communicating artery origin, posterior communicating artery, and possibly the anterior communicating artery. Currently, the recommendations are based mainly on expert opinion and guidelines, but no high-level evidence is available from clinical trials [2]. Surgical or endovascular intervention are associated certain morbidity and mortality. In a larger meta-analysis of 2460 patients, Raaymakers et al reported a mortality of 2·6% and morbidity of 10·9% [21]. Naggara et al [22] ported a meta-analysis of 7034 underwent endovascular treatment of intracranial unruptured aneurysms, 69 (1.8%) patients died. Unfavorable outcomes, including death, occurred in 4.7% of patients. In addition, there has been some concern regarding the validity of the conclusions from these meta-analyses because of issues related to publication bias [23, 24]. Findings from a larger cohort of 1917 prospectively entered and assessed patients revealed a combined morbidity and mortality at 1 year of 12·6% for those without previous haemorrhage (death, 2·7%; functional disability only, 1·4%; impaired cognitive status, 5·5%; and both functional disability and impaired cognitive status, 2·8%) and 10·1% for those with previous subarachnoid haemorrhage from some other aneurysm (death, 0·6%; functional disability only, 0·9%; impaired cognitive status, 7·1%; and both functional disability and mpaired cognitive status, 1·5%) [25]. For those patients treated surgically, morbidity and mortality was highest in those with aneurysms that were large or in the posterior circulation and in those patients older than age 50 years [25]. For these concerns, endovascular and microsurgical management of unruptured intracranial aneurysms should be undertaken in dedicated neurovascular centers to ensure the lowest complication rates [2].

Mohr *et al* [26] summarized three theories for the etiology of cerebral aneurysm. In the first theory, the aneurysm develops from congenital defects in the media of cerebral arteries; in the second theory, degenerative changes of the vessels triggered by acquired factors result in injury of the internal elastic lamina, leading to formation of the aneurysm at the area where the fragility of the vascular wall is increased locally; and in the third theory, aneurysm develops only when acquired factors are added to congenital factors. Some authors suggested a time related accumulation of degenerative changes of the cerebral arteries play an important role in the formation of aneurysms. Atherosclerotic vessel degeneration due to prolonged exposure to hypertension might affect the risk of aneurysm formation or rupture [27]. Patients with atherosclerosis also had an increased frequency of aneurysms, which corresponds with the finding that cardiovascular diseases and SAH share the risk factors smoking, hypertension, and alcohol abuse [5, 28]. The prevalence of aneurysms was very low in the first two decades of life and steadily increased after the third decade [2]. While UIAs and aneurysmal SAH are thought to be more common among elderly people [29-33]. In the population based screen Li *et al* [29] showed unruptured cerebral aneurysm significantly increase with age among 35-75 age groups.

Many publications suggested the incidence of subarachnoid hemorrhage from rupture of UIAs increases with age [5, 6, 34], from approximately 1.5-2.5 per 100,000 persons per year during the third decade of life to approximately 40-78 per 100,000 persons per year during the eighth decade [35, 36]. The risk of rupture of UIAs is about 3.05%- 5.7% per year in elderly patients, higher than in the general population [5]. Phillips *et al.* [35] reported that the annual incidence of SAH in the Framingham study was 2.8 per 100,000 for the total population, but for the elderly population, the rate was 7.8 per 100,000. Other groups noted incidence rates three times higher in the elderly population [37]. Particularly individuals aged 70 years or older is considered of high risk [38, 39, 40]. Morita suggested that unruptured cerebral aneurysms tended to burst more often in patients older than 60 years [6]. Inagawa indicated that the rupture rates of UIAs were 3 times higher in the elderly population [32].

A question of fundamental importance, which remains unanswered, is how to distinguish the growing, unhealed small aneurysm, which has the highest potential for rupture, from the stable lesion. For example, the incidence of SAH among women who smoked and had high systolic blood pressure (≥159 mmHg) was 20-fold higher than among men who had never smoked and had low systolic blood pressure (≤122 mmHg) [ 41]. The risk of rupture was higher in women and for aneurysms that were symptomatic, ≥ 10 mm, or in the posterior circulation. All patients treated conservatively should be counselled about potential risk factors for aneurysm growth and rupture. Smoking cessation should be strongly advocated. Hypertension, if present, should also be aggressively controlled. Alcohol should be used only in moderation. Clinical trials show hormone replacement therapy (HRT) seems to be associated with a reduced risk for aneurysmal SAH in post menopausal women [42,43]. Findings from several studies also suggested that cerebral aneurysm rupture risk is reduced in patients taking aspirin [44,45].Though whether the benefit of aspirin use or HRT in patients presenting with an unruptured intracranial aneurysm outweighs potential risks remain to be further studied.

In contrast to previous reports, our data demonstrated the UIA rupture rate decreased after 65 years old, both for females and males. Actually it seems that for males the UIA rupture rate was highest in early fifties, and gradually decrease therefore. Previous studies have too few patients in their old ages, while our data have sufficient cases to draw this reasonable conclusion as shown in Table 1. Our data therefore support the hypothesis that aneurysms do not grow progressively once they form but probably either rupture or stabilize and that very elderly patients are at a reduced risk of rupture compared with patients who are younger with the same-sized aneurysms [46]. As aneurysm with confirmation of aneurysm enlargement is recommend to be treated surgically or endovascularly, old patients detected with UIAs may have aggressive regular repeat MRA or CTA, though the optimal examination interval remain unknown.  It has been suggested that exposure to parental smoking in childhood is associated with an increased risk of atherosclerotic changes in adulthood [17]. It remains to be investigated whether the poor nutrient status for subjects born during 1959-1962 would have higher UIS prevalence or rupture rate.

There are a number of limitations of our study. This is a retrospective analysis of archived data. These data are from random but convenient sources, than rather than from stratified sampling; though we made efforts from obtain samples from different regions of China, and we hope the relative large number of cases would likely eliminate potential selection biases. Because we did have stratified sampling, we can only present 'relative' aneurysmal SAH incidence rate. Screening for asymptomatic cerebral aneurysms is not routinely undertaken in China. Our study simply examined ruptured aneurysms in a population with an unknown number of unruptured aneurysms. Furthermore, we do not know how many patients with ruptured aneurysms in the sampled region and sample period did not seek medical attention. It is possible that older patients with ruptured aneurysms were slightly less likely to be treated in the hospital and younger patients. Sudden deaths that result from cerebral aneurysm rupture happened outside the hospital are rarely confirmed through autopsies. Therefore, patients the worst possible outcome -death - are excluded from the final study cohort and subsequent statistical analyses. It is unknown how to best adjust the UIA rupture with the population base of China. There was a substantial demographic fluctuation during 1959-1962 period with low birth rate [13-15], this was reflected by the absolute number of UIS rupture was lower during this time. It would be a reasonable assumption on average the SAH incidence about four years ago from now. The female to male ratio has been reported to be 1.6:1 [8, 9]. The female to male ratio in this study is smaller than previously report, this is partially due to a sub-cohort which contained more males than females [11]. Overall, we expect these limitations would not have substantially affected our conclusion. Perfect studies on the epidemiology of UIAs and SAH do not exist, so we have to settle for suboptimal epidemiological evidence [3].

In conclusion, our large sample size retrospective analysis of data suggested UIA rupture rate decrease after 65 years old. Cerebral aneurysms may not grow progressively once they form but probably either rupture or stabilize and that very elderly patients are at a reduced risk of rupture compared with patients who are younger with the same-sized aneurysms. As aneurysm with confirmation of aneurysm enlargement is recommend to be treated surgically or endovascularly, patients treated conservatively have regular repeat MRA or CTA will be of paramount importance.

**Acknowledgement**: The authors thanks Dr Jian Yang of Xi'an Jiao Tong University No. 1 Hospital, Dr Zheng Han Yang of Capital Medical University Beijing Friendship Hospital, and all the co-authors of reference 11 at Wenzhou Medical University No. 1 Hospital, for providing some of the aneurysmal subarachnoid hemorrhage cases analyzed in this study, we thank Miss Yao Li for processing some of the data.